\documentclass[twocolumn,prd,aps,nofootinbib]{revtex4}
\usepackage{graphicx}
\usepackage{dcolumn}
\usepackage{bm}
\usepackage{amsmath}
\usepackage{epsfig}
\usepackage{graphicx}
\usepackage{hyperref}
\usepackage{url}
\usepackage{multirow}

\def\be{\begin{equation}}
\def\ee{\end{equation}}
\def\bea{\begin{eqnarray}}
\def\eea{\end{eqnarray}}
\newcommand{\Mpc}{\text{Mpc}}
\newcommand{\kunit}{~\text{h}~\Mpc^{-1}}
\newcommand{\lunit}{~\text{h}^{-1}~\Mpc}

\begin{document}

\title{Searching for dark matter isocurvature initial conditions with N-body Simulations}

\author{Jie Liu$^{a}$}
\email{liujie@ihep.ac.cn}

 \affiliation{${}^a$Institute of High
Energy Physics, Chinese Academy of Science,
P.O. Box 918-4, Beijing 100049, P. R. China}

\begin{abstract}
Small fraction of isocurvature perturbations may exist and correlate
with adiabatic perturbations in the primordial perturbations.
Naively switching off isocurvature perturbations may lead to
biased results.  We study the effect of dark matter isocurvature
on the structure formation through N-body simulations.  From the best
fit values, we run four sets of simulation with different initial
conditions and different  box sizes. We find that, if the fraction
of dark matter isocurvature is small, we can not detect its signal
through matter  power spectrum and two point correlation function with
large scale  survey. However, the halo mass function can give an obvious
signal. Compared to $5\%$ difference on matter power spectrum, it can get 
$37\%$ at $z = 3$ on halo mass function. This indicates that future high 
precise cluster count experiment can give stringent constraints on dark 
matter isocurvature perturbations.
\end{abstract}

\maketitle

\section{introduction}

Recent high accuracy observations, such as  cosmic microwave background
radiation (CMB)\cite{wmap7}, and the large scale structure (LSS)\cite{sdss},
provide us wealthy information on the universe. Currently, so called
concordance cosmological model, in which approximately scale-invariant,
Gaussian, adiabatic  primordial perturbations seed the structure of our 
universe, is mostly favored.  However, there exist  some models which predict
non-negligible  isocurvature perturbations,  such as multiple scalar
fields inflation\cite{Liddle:1998jc,  Kanti:1999vt, Copeland:1999cs,
Langlois:1999dw, Wands:2002bn, Piao:2002vf, Dimopoulos:2005ac,
 Langlois:2008wt, Cai:2009hw, Cai:2008if, Cai:2010wt} and
curvaton\cite{Lyth:2001nq,Moroi:2001ct,  Lyth:2002my} models.
The primordial isocurvature perturbations then can induce the cosmological
matter  perturbations such as cold dark matter (CDM), baryons, dark energy
and neutrino\cite{Seckel:1985tj,  Linde:1991km, Linde:1996gt,Liu:2010ba} in 
the radiation era.

Although the pure isocurvature perturbations have already been ruled out by the
observation of Boomerang and MAXIMA-1\cite{Enqvist:2000hp}, models  with
primordial perturbations comprised of dominate adiabatic and a small fraction of 
isocurvature modes\cite{Gordon:2002gv, Crotty:2003rz,  Bucher:2004an,
Moodley:2004nz, KurkiSuonio:2004mn, Beltran:2005xd, Bean:2006qz, Trotta:2006ww,
Sollom:2009vd,  Valiviita:2009bp, Beltran:2005gr,  Mangilli:2010ut, Li:2010yb} still 
survive. Due to the quality of data, we still  can not confirm  the  destiny of the
 isocurvature perturbation.  Moreover, if we  switch off the isocurvature perturbations
naively, the result would be misleading\cite{Zunckel:2010mm, Liu:2010ba}.
So, it is important to seek a way  to detect the isocurvature perturbations
and give stringent constraints on isocurvature perturbations parameters.

In  Ref. \cite{Li:2010yb}, we have given constraints on parameters of two different
cosmological models  with latest astronomical data.  One is with pure adiabatic
initial condition (IC) and the other is with mixed IC in which  small fraction of
isocurvature mode is correlated with dominating adiabatic mode through $\cos\Delta$.
We found that compared to adiabatic mode, the isocurvature perturbations have
smaller amplitude ($A_s^{iso}/A_s^{adi} \sim 10^{-2}$) and bluer tilt
($n_s^{iso} \sim 2$). That is to say the isocurvature modes is pronounced
on small scales ($k > 1\kunit$). However, on these scales, the structure grows non-linearly, 
so we can not study the details of evolution in a full analytical
way and then test them against future high-precision observations.

On the other way, N-body simulations plays more and more important role in 
cosmology nowadays, especially in the study of nonlinearity. It can make predictions
and compare with observations for a specific model. It can also be used to check 
the validity of a particular  method \cite{Bagla:2004au}. 

	In this work, we study the effects of CDM isocurvature perturbations on LSS  by
	implementing N-body simulations. Given the best fit parameters in Ref. \cite{Li:2010yb},
	we carry out four sets of simulations with different ICs in different boxes.
	We find that,  unlike the power spectrum and two point correlation function,
	mass function is sensitive to dark matter isocurvature perturbations.
	This indicates  that, we can give stringent constraints on dark matter
	isocurvature perturbations  with future high  precision cluster counts experiment.

	The structure of this paper is organized as follows. In Sec. \ref{sec:ic},
	we  briefly introduce the CDM isocurvature perturbation and set the ICs
	for N-body simulation. In Sec. \ref{sec:nbody},  we describe the details of
	our N-body simulation and present the result. We give summary
	 in Sec. \ref{sec:summary}.

\section{Correlated adiabatic and isocurvature perturbations}
\label{sec:ic}
Generally, to calculate matter power spectra $P(k)$ numerically, one start
from the time $t_{ic}$ deep in the radiation dominated era,  when all interesting
scales of perturbations are outside the horizon. However, $t_{ic}$ is
different  from the time $t_*$  when the corresponding mode $k$ exits horizon
during inflation. Therefore,  One need  transfer function $T_{ij}$ to transform 
perturbations from  $t_*$ to $t_{ic}$.

	As we know,  in the absence of isocurvature perturbations, the adiabatic
	perturbations  $\mathcal{R}$ are conserved on superhorizon scales. 
	On the contrary, the isocurvature perturbations $\mathcal{S}$ can evolve on
	superhorizon scale and can also seed adiabatic perturbations. Thus, the transfer
	function $T_{ij}$ can be written as\cite{Amendola:2001ni}
	\bea\label{evol}
	\begin{bmatrix}\mathcal{R}(t_{ic}) \\ \mathcal{S}(t_{ic}) \end{bmatrix} =
	\begin{bmatrix} 1 & T_\mathcal{RS}\\0 & T_\mathcal{SS} \end{bmatrix}
	\begin{bmatrix} \mathcal{R}(t_*) \\ \mathcal{S}(t_*)\end{bmatrix},
	\eea
where $T_{ij}$ is model dependent. To investigate isocurvature
perturbation without making use of any specific model, one  often parametrizes the
transfer function in the simple form of power law.

For the Gaussian statistics, which is predicted by inflation, the power spectra
characterize all the information.
	\be
	\mathcal{P}_{ij} \equiv \frac{k^3}{2\pi^2}\langle\mathcal{X}_i({\bf k})
	    \mathcal{X}_j({\bf k}')\rangle\delta({\bf k}-{\bf k}'),
	\ee
where $\mathcal{X}_1 = \mathcal{R}$ and $\mathcal{X}_2 = \mathcal{S}$.
We can parametrize primordial power spectra $\mathcal{P}(k)$ at $t_{ic}$ as
\be\label{primordial_pk}\mathcal{P}^{ij}(k)=A^{ij}_s(\frac{k}{k_0})^{n^{ij}_s-1},\ee
where $k_0$ is pivot scale, and both $A_s^{ij}$ and $n^{ij}_s$ are $2$ dimensional  symmetric
matrices  which denote the amplitude and power index, respectively. The amplitude of the 
spectra $A_s^{ij}$  can be  written as
	 \bea A_s^{ij}=\begin{pmatrix}
	A_s^{\rm adi}&\sqrt{A_s^{\rm adi}A_s^{\rm iso}}\cos\Delta\\
	 \sqrt{A_s^{\rm adi}A_s^{\rm iso}}\cos\Delta & A_s^{\rm iso}
	\end{pmatrix},
	 \eea where $\cos\Delta=A_s^{\rm adi,iso}/\sqrt{A_s^{\rm
	adi}A_s^{\rm iso}}$ describes the correlation between adiabatic mode and isocurvature mode
	\cite{Langlois:1999dw}, $A_s^{\rm adi}$ and
	$A_s^{\rm iso}$ stand for the amplitude of adiabatic and isocurvature modes,
	respectively. The spectra index $n^{ij}_s$ is
	\bea n_s^{ij}=\begin{pmatrix}
	n_s^{\rm adi}&n^{\rm cor}_s\\
	 n^{\rm cor}_s & n_s^{\rm iso}
	\end{pmatrix},
	 \eea
	with $n^{\rm adi}_s$ and $n^{\rm iso}_s$ being the spectra indices for adiabatic
	and isocurvaure modes. Here, we have used the approximation $n^{\rm cor}_s=\frac{n_s^{11}
	+n_s^{22}}{2}$\cite{KurkiSuonio:2004mn} for simplicity.

	Since both adiabatic and isocurvature perturbations seed the large scale structure as
	\be
	\delta = \delta_{adi}+\delta_{iso}
	\ee
	Then  besides the normal adiabatic term, two other terms, isocurvature
	 and cross-correlation terms, emerge in the expression of matter  power spectrum,
	 \bea P(k)&=&A_s^{\rm adi}\hat{P}^{\rm adi}(k)+A_s^{\rm iso}\hat{P}^{\rm
	iso}(k)\nonumber\\
	 && +2\sqrt{A_s^{\rm adi}A_s^{\rm iso}}\cos\Delta\hat{P}^{\rm
	adi,iso}(k), \label{pk}\eea
	where $\hat{P}^{\rm i}(k)$ can be described as
	\be
	\hat{P^{ij}}(k) = (\frac{k}{k_0})^{n_s^{ij}-1}T^i(k)T^j(k),
	\ee
	with $T^i(k)$ being transfer function of matter for  IC $i$.

\begin{figure}[htbp]
	\centering
	\includegraphics[width=0.85\linewidth]{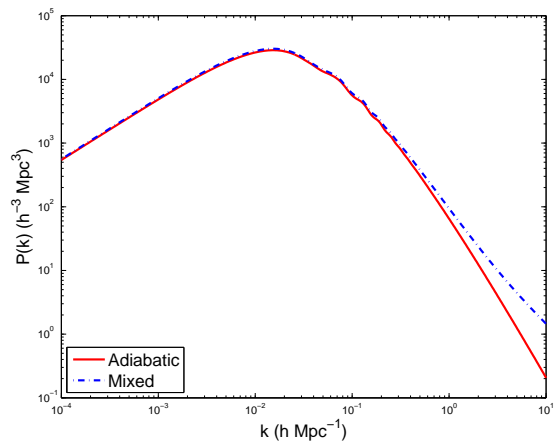}
	\caption{Linear matter power spectra for the models in Table \ref{tab:cons}. The red
	solid line corresponds to standard ${\Lambda}CDM$ model with adiabatic initial condition
	while the blue dash-dotted line is given by mixed initial condition.}\label{fig:pkth}
	\end{figure}

	We have given constraints on these parameters with latest observations
	in Ref. \cite{Li:2010yb}. The the best fit values are listed  in Table 	\ref{tab:cons}.  
Using CAMB\cite{camb}, We  also sketch the power spectra
	in Fig. \ref{fig:pkth} with best fit values.

	\begin{table}
	\centering
	\caption{The best fit value for models with different initial conditions.}
	\begin{tabular}{c|c|c}
	\hline
	\hline
	Parameters	& Aidabatic	 &	Mixed\\
	\hline
	$\Omega_b$ &	$0.046$	& $0.044$ \\
	$\Omega_m$ & $0.280$ & $0.267$ \\
	$\Omega_{\Lambda}$ & $0.720$ & $0.733$\\
	$h$ & $0.700$ & $0.714$\\
	$10^9A_s^{adi}$ & $2.176$ & $2.420$ \\
	$n_s^{adi}$ &$0.960$& $0.965$\\
	$10^{10}A_s^{iso}$ & $- $& $0.081$\\
	$n_s^{iso}$& $-$ & $2.716$\\
	$\cos\Delta$ & $-$ & $0.173$ \\
	$\sigma_8$ & $0.820$  &  $0.865$ \\
	\hline
	\end{tabular}
	\label{tab:cons}
	\end{table}

	\section{N-body Simulation}
	\label{sec:nbody}

	We  perform our simulations with GADGET-2\footnote{Available at
	\url{http://www.mpa-garching.mpg.de/gadget/}}\cite{gadget}, a massively parallel TreePM-SPH
	(Tree Particle Mesh-Smoothed Particle Hydrodynamics) code. For collisionless particles,
	the gravitational field is calculated with a low-resolution particle-mesh(PM)
	algorithm on large scales,  while  forces are delivered by tree on small scales.
	We do not use the SPH part since only cold dark matter particles are considered in this
	work.

	\subsection{Initial Conditions}
	With power spectrum plotted in Fig. \ref{fig:pkth}, we can generate
	positions and velocity ICs for particles at cosmic time $\tau$
	using the Zel'dovich Approximations(ZA).
	\bea
	{\bf x}({\bf q},\tau) &=& {\bf q} +D^+(\tau){\bf \Psi}({\bf q}), \\
	{\bf v}({\bf q},\tau) &=& \dot D^+(\tau){\bf\Psi}({\bf q}),
	\eea
	where ${\bf x}$ is the perturbed comoving coordinates and
	${\bf v}\equiv\frac{d{\bf x}}{d{\bf\tau}}$ is the proper peculiar
	velocity. {\bf q}, the lagrangian coordinates generated from
	glass configuration\cite{White:1994bn},  denote the unperturbed
	comoving position. $D^+(\tau)$ is the linear growth factor normalized
	to $z=0$.  ${\bf\Psi}({\bf q})$ is displacement field calculated from
	the density fluctuation field which is the convolution of a random white
	noise with the square root of the linear power spectrum\cite{Sirko:2005uz}.

	The ICs are set at $z = 49$ when the second order Lagrangian perturbations
	correction can be ignored safely.  we run  four sets of simulations with different 
        box sizes to explore the differences between two initial 
	conditions on different scales. The larger boxes whose length is $1000\lunit$ provide 
	good statistics  on large scales from $k\sim 10^{-3}\kunit$ to $k\sim 1\kunit$, while the smaller 
	boxes with $L= 100 \lunit$ can give high resolution extending to $k \sim 10 \kunit$.  
	In Set I, the mass resolution is $5.8\times10^{11}~{\text h}^{-1}M_\odot$ with $N_{p} = 512^3$
	while in Set II the mass resolution is about $2.4\times10^{9}~{\text h}^{-1}M_\odot$.
	The force resolution  is taken as $\sim0.5\%$ of the mean particle
	interval (Tab. \ref{tab:sim}). 

	Because we can get only one value of $\sigma_8$ for a specific survey and to cease the effect 
	of different choices of cosmological parameters, we renormalize the power spectra in Fig. \ref{fig:pkth} to the
        same  $\sigma_8 = 0.8$. This setting may make results present below  not so obvious, however, 
	what we are interested in is the relative difference, which is independent on the renormalization.

	\begin{table}
	\centering
	\caption{Simulation details for adiabatic and mixed initial conditions.}
	\begin{tabular}{c|c|c|c|c}
	\hline\hline
	\label{tab:sim}
	 &\multicolumn{2}{c|}{Set I} &\multicolumn{2}{c}{Set II} \\
	 \hline
	 Initial Condition & adiabatic & Mixed & adiabatic & Mixed \\
	 \hline
	 $L_{box}(h^{-1} Mpc)$ & $1000$  & $1000$  &  $100$ & $100$ \\
	 $N_{part}$ & $512^3$ & $512^3$ & $320^3$ & $320^3$ \\
	 $L_{soft}(h^{-1} kpc)$ & $10$  & $10$ & $5$ & $5$ \\
	 $z_{start}$  & $49$ & $49$ & $49$ & $49$ \\
	 \hline
	\end{tabular}
	\end{table}

	\subsection{Numerical Results}
	\subsubsection{Correlation Function}
	The baryon acoustic oscillation (BAO), as a standard ruler, is a powerful
	tool to study the dark energy. It is also a useful tool to detect the dark
	matter isocurvature perturbation. The  presence of dark matter isocurvature
	perturbation would alter the position of first peak in the CMB angular power
	spectrum, which is the right scale of BAO.  The peak in the two point
	correlation function and the wiggles in the power spectrum are the useful
	tools to track the behavior of BAO\cite{Eisenstein:2005su}.

\begin{figure}[htbp]
	\centering
	\includegraphics[width=0.85\linewidth]{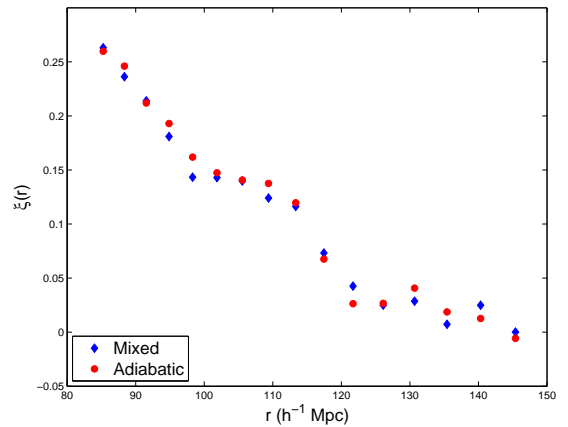}
	\caption{The 2-point correlation function for two simulations in the large box at $z=0$.
	The blue stars denote mixed case while the red points are for adiabatic IC.
	}\label{fig:2pcf}
	\end{figure}

	We calculate the 2-point correlation function for Set I at $z=0$
	with the pair-count estimator proposed by  Landy $\&$  Szalay\cite{Landy:1993yu}:
	\be
	\xi(r) = \frac{DD-2DR+RR}{RR},
	\ee
	where $DD$ and $RR$ are the autocorrelation function of the simulation particles and
	randomly sampled points respectively, $DR$ is the cross-correlation between the data
	and random points. From Fig. \ref{fig:2pcf}, we can find that,  the position and width
	of BAO from which $H(z)$ and $D_A(Z)$ are extracted, are almost the same for the two
	different simulations.  This result is reasonable in two aspects. Firstly, the BAO observation 
	 mainly depends on the background parameter, while has little to do with the origin of the perturbations;
	secondly,  we set ICs with best fit parameters. The behavior on large scales ($k < 0.2\kunit$) 
	is well constrained\cite{sdssdr7}, especially
	the position of first peak in the CMB angular power spectrum\cite{wmap7}.
	It implies that, if the isocurvature fraction is small enough,
	we can  not discriminate  two initial conditions from BAO observation.  This is
	an important systematic error in constraining dark energy from BAO  data\cite{Mangilli:2010ut,
	Zunckel:2010mm}.

	\subsubsection{Power Spectra}
	Power spectrum, defined as the Fourier transformation of two point correlation function,
	is the key physical quantity in  understanding clustering properties.
	With a Gaussian initial condition as selected in this paper,
	power spectrum  gives a complete statistical description of fluctuations.

\begin{figure}[htbp]
	\centering
	\includegraphics[width=0.85\linewidth]{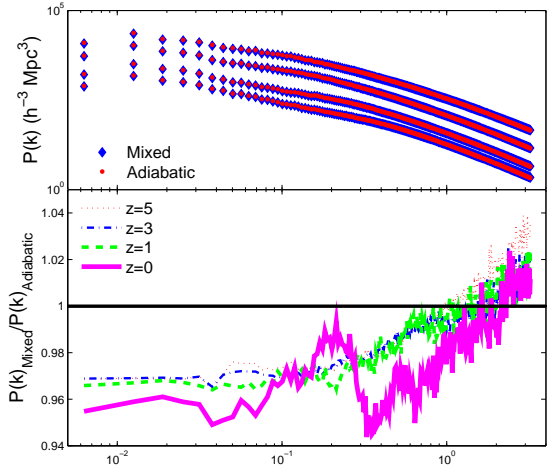}
	\caption{Top panel:The power spectrum of simulation Set I. From bottom
	to top: $z=5$, $z=3$, $z=1$, $z=0$; The blue stars stand for mixed initial condition
	while red points denote adiabatic $\Lambda$CDM. Bottom panel: the ratio of matter power spectrum between
	two initial conditions. The thickness is proportional to the scale factor.}\label{fig:power512}
	\end{figure}

	`POWMES' is a power spectrum estimator based on the Taylor expansion of the trigonometric
	functions\cite{Colombi:2008dw}.  The further `foldings' scheme makes it possible to  give
	an accuracy measurement of power spectrum up to a scale
	\be
	k_{max} = k_{ny}\times2^{n_{fold}-1},
	\ee
	where $k_{ny}=\frac{2\pi}{L_{box}}\frac{N_p}{2}$ is the Nyquist frequency and $n_{fold}$  is
	the number of foldings which is set as $2$ in this work.
	We plot the power spectra at different redshifts as well as their ratios of
	Set I in Fig. \ref{fig:power512}.

	From Fig. \ref{fig:power512}, we can find that power spectra of simulation with mixed
	IC is smaller than the one in adiabatic case on large scales (small $k$).  $r$,
	the ratio of $P_{mixed}$ to $P_{adiabatic}$, grows as a function of time
	on large scales, $k > 2\kunit$ and decreases on small scales, $k < 2\kunit$.
	For $k \sim 0.02\kunit$,  $r$ reaches about  $0.95$  at $z = 0$. That is to say,
	 the largest discrepancy  in power spectrum is about $5\%$.

	What we have to keep in mind is that we have renormalized the initial power spectra
	to the same $\sigma_8$ and the weight  $k^2(\frac{\sin(kr)}{kr^3}-\frac{\cos(kr)}{kr^2})^2$
	peaks around $k \sim 0.3\kunit$.  So,  the power spectra we got from N-body simulations
	with mixed IC should be similar to the one with adiabatic IC on large scales while be larger
	on small scales without renormalization.

	\subsubsection{Halo Mass Function}
	Mass function is defined as the abundance of dark matter haloes in a specific mass 
	ranges.  It is a key quantity to describe the large scale structure in the nonlinear
	regime. Press and  Schechter firstly provided the theoretical description in a simple
	spherical collapse model\cite{Press:1973iz}. Subsequently, people made some improvements
	on this simple modelling and introduced more complex ellipsoidal collapse models\cite{Sheth:1999su}.
	Meanwhile,  a lot of literatures try to fit the halo mass function in the manner 
	of N-body simulation\cite{Sheth:1999mn, Jenkins:2000bv,  Warren:2005ey, Tinker:2008ff}.

	We identify  haloes with AHF\footnote{Available at \url{http://popia.ft.uam.es/AMIGA}
	} (Amiga's Halo Find)\cite{AHF}, an adaptive mesh based finder. 
	After placing grid across the box and further refinement, AHF  assigns each particle to 
	a grid with cloud-in-cell (CIC) interpolation.  Then, AHF probes the halo at each density 
	peak using spherical overdensity (SO) algorithm. The radius of sphere is grown until 
	the interior density  reaches a specific value
	\be
	\Delta\equiv\frac{M_{\Delta}}{4/3\pi R_{\Delta}^3\rho_{bkg}},
	\ee
	where $\rho_{bkg}\equiv \Omega_m \rho_{crit} (1+z)^3$ is the mean density of whole box.
	To compare mass functions of these two different models, we set $\Delta$ as $200$ in this paper.
	With these scheme, AHF can find all structures and substructures simultaneously. 
	Moreover, we only keep haloes with at least 20 dark  matter particles, i.e. the mimimum
	mass of haloes is around  $4\times10^{10} h^{-1} M_\odot$.

	We introduce the  cumulative mass function which is defined as mean number
	densities of haloes with mass larger than a specific mass,
	\be
		n(>M) = \frac{N(>M)}{L_{box}^3},
	\ee
where $N(>M)$ is the number of haloes with mass greater than M. 
This is related to the  mass function $f(\sigma)$ through
\bea
n(>m)  &=& \int_M^\infty \frac{dn}{dM}dM \nonumber\\
	&=& \int_M^\infty f(\sigma)\frac{\bar\rho_m(z=0)}{M}\frac{d\ln\sigma^{-1}}{dM}dM
\eea

\begin{figure*}[htbp]
\centering
\includegraphics[width=0.9\linewidth]{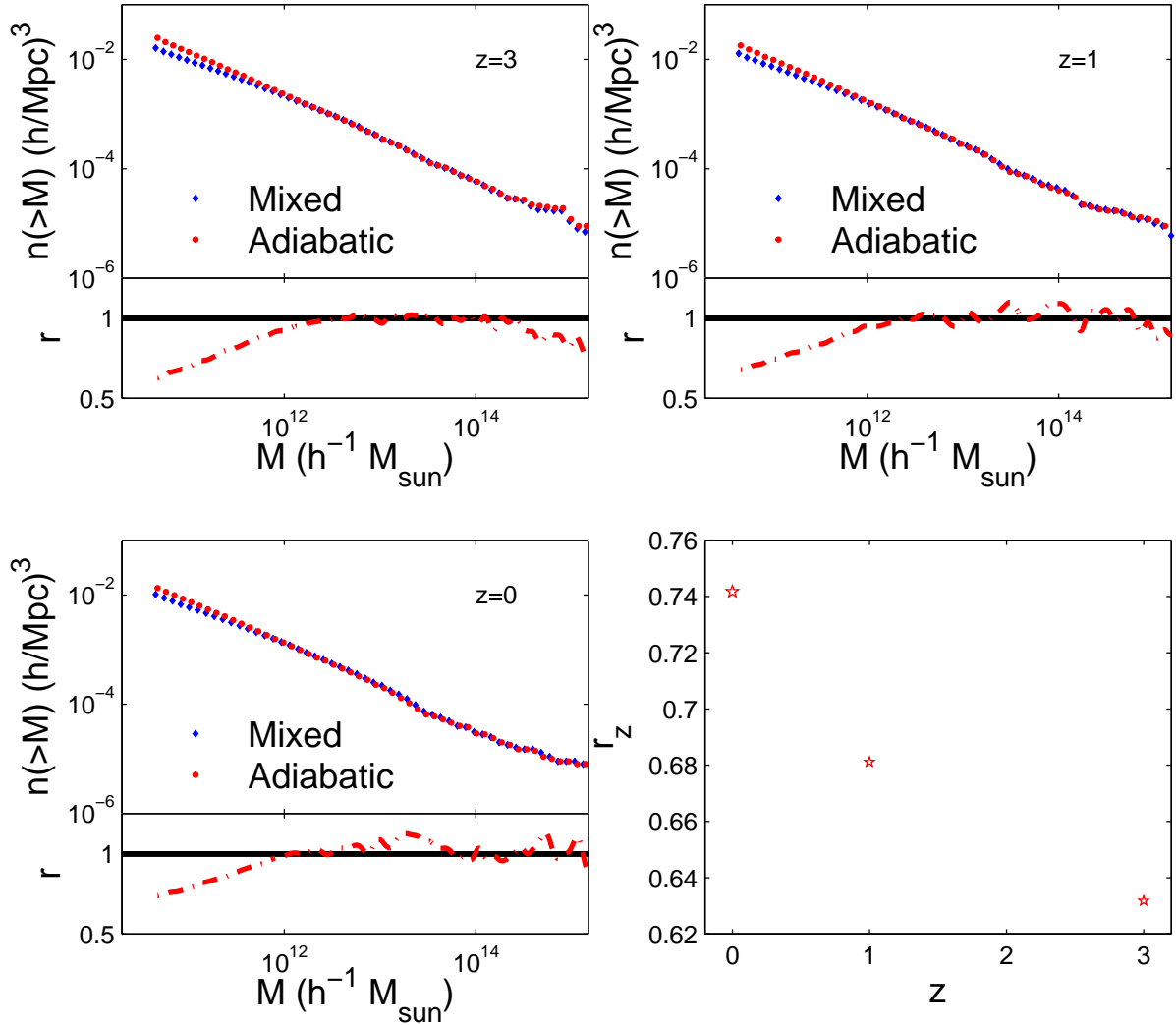}
\caption{Cumulative mass function  as well as the ratio of two situations for simulation Set II. Top left are for
$z=3$, top right $z=1$ and bottom left $z=0$.  The bottom right is the plot of  discrepancy evolvement at $M = 
5\times10^{10} M_\odot$. }
\label{fig:mf320}
\end{figure*}

The cumulative  mass function as well as the ratio $r\equiv{n_{mixed}}/{n_{adia}}$ for
Simulation Set II at redshifts $z = 3$, $1$, $0$ are sketched in 
Fig. \ref{fig:mf320}.  We can find that the mass function is almost the same on large mass
scale, from $10^{12} M_{\odot}$ to $10^{15}M_{\odot}$.  However, there are more haloes for adiabatic IC than for 
mixed IC on small mass scale ( $M < 10^{12} M_\odot$ ). Moreover, the discrepancy increases as the mass gets smaller. 
For $M\sim5\times10^{11}M_\odot$,  the difference is about $26\%$ at $z=0$.  We also plot the ratio $r$ 
against redshift $z$ for $M = 5\times10^{10} h^{-1} M_\odot$ in the bottom right of Fig. \ref{fig:mf320}.  
We find that the ratio decreases as time evolves. Compared to $r = 74\%$ at $z = 0$, the ratio at $z = 3$ is
only $63\%$. This behavior can be ascribed to the late-time non-linear evolution. That is to say, a high-redshift
survey is helpful  to seek the DM isocurvature perturbations signal. 

Since we have renormalized to the same $\sigma_8$, the power on large scales for mixed IC is less than for
adiabatic IC (Fig. \ref{fig:power512}),  and there should be more haloes for mixed IC than for the adiabatic one without
renormalization.

\section{Summary}
\label{sec:summary}
Isocurvature perturbations, inevitably generated from multi-field  inflation  or curvaton models, 
can be used to test these models.  Although pure isocurvature perturbation models have already been 
ruled out,  there still exists possibility that a  small fraction of isocurvature mode is correlated 
to  the dominating adiabatic perturbation.  This is important to the parameter estimation, since rough 
ignorance would lead to biased result.

With the best fit values obtained in Ref.\cite{Li:2010yb},   we perform four sets of N-body simulations
to seek a best way to detect the isocurvature perturbations.  We find that, if the fraction is small 
enough($A_{iso}/A_{adia}\sim 3\%$)  we can not peek it in the BAO  observation. The position and the
width of the bump in two-point  correlation function,  which  mainly depend on the background parameter,
are almost the same.  There are some differences in matter power spectrum and halo mass function. However,
the  $5\%$ difference in matter power  spectra  makes it hard to be observed.  On the contrary,  the deviation in
the halo mass function is obvious. The difference is getting larger as we go to  higher redshift.  
For $M \sim 5\times10^{10} M_\odot$ the discrepancy can get $37\%$ at $z = 3$.   This implies that, with future precise 
cluster number count observations,  we can detect the initial condition of dark matter isocurvature and give stringent
constraints.

\section*{Acknowledgements}
We perform our simulations on  Deepcomp7000 of Supercomputing Center, Computer Network Information
Center of Chinese Academy of Sciences. We thank Shi Shao, Wenting Wang, Jia-Xin Han, Hong Li, Jun-Qing Xia, Yi-Fu Cai
 Taotao Qiu and Mingzhe Li for helpful discussions.


\begin{thebibliography}{nn}
\bibitem{wmap7}
  E.~Komatsu {\it et al.}  [WMAP Collaboration],
  Astrophys.\ J.\ Suppl.\  {\bf 192}, 18 (2011).

    \bibitem{sdss}
  B.~A.~Reid {\it et al.},
  Mon.\ Not.\ Roy.\ Astron.\ Soc.\  {\bf 404}, 60 (2010).

  \bibitem{Liddle:1998jc}
  A.~R.~Liddle, A.~Mazumdar, F.~E.~Schunck,
  Phys.\ Rev.\  {\bf D58}, 061301 (1998).

\bibitem{Kanti:1999vt}
  P.~Kanti, K.~A.~Olive,
  Phys.\ Rev.\  {\bf D60}, 043502 (1999).

\bibitem{Copeland:1999cs}
  E.~J.~Copeland, A.~Mazumdar, N.~J.~Nunes,
  Phys.\ Rev.\  {\bf D60}, 083506 (1999).

\bibitem{Langlois:1999dw}
  D.~Langlois,
  Phys.\ Rev.\  {\bf D59}, 123512 (1999).

  \bibitem{Wands:2002bn}
  D.~Wands, N.~Bartolo, S.~Matarrese and A.~Riotto,
  Phys.\ Rev.\  D {\bf 66}, 043520 (2002).


\bibitem{Piao:2002vf}
  Y.~S.~Piao, R.~G.~Cai, X.~m.~Zhang and Y.~Z.~Zhang,
  Phys.\ Rev.\  D {\bf 66}, 121301 (2002).

\bibitem{Dimopoulos:2005ac}
  S.~Dimopoulos, S.~Kachru, J.~McGreevy and J.~G.~Wacker,
  JCAP {\bf 0808}, 003 (2008).

  \bibitem{Langlois:2008wt}
  D.~Langlois, S.~Renaux-Petel, D.~A.~Steer and T.~Tanaka,
  Phys.\ Rev.\ Lett.\  {\bf 101}, 061301 (2008).


\bibitem{Cai:2009hw}
  Y.~-F.~Cai, H.~-Y.~Xia,
  Phys.\ Lett.\  {\bf B677}, 226-234 (2009).


\bibitem{Cai:2008if}
  Y.~-F.~Cai, W.~Xue,
  Phys.\ Lett.\  {\bf B680}, 395-398 (2009).

\bibitem{Cai:2010wt}
  Y.~-F.~Cai, J.~B.~Dent, D.~A.~Easson,
  [arXiv:1011.4074 [hep-th]].

\bibitem{Lyth:2001nq}
  D.~H.~Lyth and D.~Wands,
  Phys.\ Lett.\  B {\bf 524}, 5 (2002).

  \bibitem{Moroi:2001ct}
  T.~Moroi and T.~Takahashi,
  Phys.\ Lett.\  B {\bf 522}, 215 (2001)
  [Erratum-ibid.\  B {\bf 539}, 303 (2002)].

\bibitem{Lyth:2002my}
  D.~H.~Lyth, C.~Ungarelli and D.~Wands,
  Phys.\ Rev.\  D {\bf 67} (2003) 023503.

\bibitem{Seckel:1985tj}
  D.~Seckel and M.~S.~Turner,
  Phys.\ Rev.\  D {\bf 32}, 3178 (1985).

\bibitem{Linde:1991km}
  A.~D.~Linde,
  Phys.\ Lett.\  B {\bf 259}, 38 (1991).

   \bibitem{Linde:1996gt}
  A.~D.~Linde and V.~F.~Mukhanov,
  Phys.\ Rev.\  D {\bf 56}, 535 (1997).

\bibitem{Liu:2010ba}
  J.~Liu, M.~Li and X.~Zhang,
  arXiv:1011.6146 [astro-ph.CO].


  \bibitem{Enqvist:2000hp}
  K.~Enqvist, H.~Kurki-Suonio and J.~Valiviita,
  Phys.\ Rev.\  D {\bf 62}, 103003 (2000).

  \bibitem{Gordon:2002gv}
  C.~Gordon, A.~Lewis,
  Phys.\ Rev.\  {\bf D67}, 123513 (2003).


  \bibitem{Crotty:2003rz}
  P.~Crotty, J.~Garcia-Bellido, J.~Lesgourgues {\it et al.},
  Phys.\ Rev.\ Lett.\  {\bf 91}, 171301 (2003).

  \bibitem{Bucher:2004an}
  M.~Bucher, J.~Dunkley, P.~G.~Ferreira {\it et al.},
  Phys.\ Rev.\ Lett.\  {\bf 93}, 081301 (2004).

  \bibitem{Moodley:2004nz}
  K.~Moodley, M.~Bucher, J.~Dunkley {\it et al.},
  Phys.\ Rev.\  {\bf D70}, 103520 (2004).

  \bibitem{KurkiSuonio:2004mn}
  H.~Kurki-Suonio, V.~Muhonen, J.~Valiviita,
  Phys.\ Rev.\  {\bf D71}, 063005 (2005).

 \bibitem{Beltran:2005xd}
  M.~Beltran, J.~Garcia-Bellido, J.~Lesgourgues {\it et al.},
  Phys.\ Rev.\  {\bf D71}, 063532 (2005).

  \bibitem{Bean:2006qz}
  R.~Bean, J.~Dunkley, E.~Pierpaoli,
  Phys.\ Rev.\  {\bf D74}, 063503 (2006).

\bibitem{Trotta:2006ww}
  R.~Trotta,
  Mon.\ Not.\ Roy.\ Astron.\ Soc.\  {\bf 375}, L26 (2007).

\bibitem{Sollom:2009vd}
  I.~Sollom, A.~Challinor and M.~P.~Hobson,
  Phys.\ Rev.\  D {\bf 79}, 123521 (2009).


\bibitem{Valiviita:2009bp}
  J.~Valiviita and T.~Giannantonio,
  Phys.\ Rev.\  D {\bf 80}, 123516 (2009).

  \bibitem{Beltran:2005gr}
  M.~Beltran, J.~Garcia-Bellido, J.~Lesgourgues and M.~Viel,
  Phys.\ Rev.\  D {\bf 72}, 103515 (2005).

\bibitem{Mangilli:2010ut}
  A.~Mangilli, L.~Verde and M.~Beltran,
  JCAP {\bf 1010}, 009 (2010).

  \bibitem{Li:2010yb}
  H.~Li, J.~Liu, J.~Q.~Xia and Y.~F.~Cai,
  arXiv:1012.2511 [astro-ph.CO].


    \bibitem{Zunckel:2010mm}
  C.~Zunckel, P.~Okouma, S.~M.~Kasanda, K.~Moodley and B.~A.~Bassett,
  Phys.\ Lett.\  B {\bf 696}, 433 (2011).


\bibitem{Bagla:2004au}
  J.~S.~Bagla,
  Curr.\ Sci.\  {\bf 88}, 1088 (2005).

\bibitem{Amendola:2001ni}
  L.~Amendola, C.~Gordon, D.~Wands and M.~Sasaki,
  Phys.\ Rev.\ Lett.\  {\bf 88}, 211302 (2002).

  \bibitem{camb}
  A.~Lewis and S.~Bridle,
  Phys.\ Rev.\  D {\bf 66}, 103511 (2002).

  \bibitem{gadget}
  V.~Springel,
  Mon.\ Not.\ Roy.\ Astron.\ Soc.\  {\bf 364}, 1105 (2005).


\bibitem{White:1994bn}
  S.~D.~M.~White,
  arXiv:astro-ph/9410043.

  \bibitem{Sirko:2005uz}
  E.~Sirko,
  Astrophys.\ J.\  {\bf 634}, 728 (2005).

  \bibitem{Eisenstein:2005su}
  D.~J.~Eisenstein {\it et al.}  [SDSS Collaboration],
  Astrophys.\ J.\  {\bf 633}, 560 (2005).

 \bibitem{Landy:1993yu}
  S.~D.~Landy and A.~S.~Szalay,
  Astrophys.\ J.\  {\bf 412}, 64 (1993).

  \bibitem{sdssdr7}
  B.~A.~Reid {\it et al.},
  Mon.\ Not.\ Roy.\ Astron.\ Soc.\  {\bf 404}, 60 (2010).

\bibitem{Colombi:2008dw}
  S.~Colombi, A.~H.~Jaffe, D.~Novikov and C.~Pichon,
  arXiv:0811.0313 [astro-ph].

  \bibitem{AHF}
  S.~R.~Knollmann and A.~Knebe,
  Astrophys.\ J.\ Suppl.\  {\bf 182}, 608 (2009).

\bibitem{Press:1973iz}
  W.~H.~Press and P.~Schechter,
  Astrophys.\ J.\  {\bf 187} (1974) 425.

\bibitem{Sheth:1999su}
  R.~K.~Sheth, H.~J.~Mo and G.~Tormen,
  Mon.\ Not.\ Roy.\ Astron.\ Soc.\  {\bf 323}, 1 (2001).

\bibitem{Sheth:1999mn}
  R.~K.~Sheth and G.~Tormen,
  Mon.\ Not.\ Roy.\ Astron.\ Soc.\  {\bf 308}, 119 (1999).

\bibitem{Jenkins:2000bv}
  A.~Jenkins {\it et al.},
  Mon.\ Not.\ Roy.\ Astron.\ Soc.\  {\bf 321}, 372 (2001).

\bibitem{Warren:2005ey}
  M.~S.~Warren, K.~Abazajian, D.~E.~Holz and L.~Teodoro,
  Astrophys.\ J.\  {\bf 646}, 881 (2006).

\bibitem{Tinker:2008ff}
 J.~L.~Tinker {\it et al.},
  Astrophys.\ J.\  {\bf 688}, 709 (2008).
\end{thebibliography}
\end{document}